# Empowering Communication: Speech Technology for Indian and Western Accents through AI-powered Speech Synthesis

Vinotha R [1], Hepsiba D [2*], L. D. Vijay Anand [3], Deepak John Reji [4]

[1] Department of Robotics Engineering

[1] vinothar@karunya.edu.in

[2] Department of Biomedical Engineering

[2] hepsiba@karunya.edu

[3] Department of Robotics Engineering

[3] vijayanand@karunya.edu

[4] Environmental Resources Management

[4] deepak.reji@erm.com

**Abstract**

Neural Text-to-speech (TTS) synthesis is a powerful technology that can generate speech using neural networks. One of the most remarkable features of TTS synthesis is its capability to produce speech in the voice of different speakers. This paper builds upon the foundation of neural TTS synthesis, particularly focusing on voice cloning and speech synthesis capabilities for Indian accents. This stands in contrast to most existing systems, which are predominantly trained on Western accents. First, a LSTM based speaker verification system identifies distinctive speaker traits. Next, a synthesizer, acting as a sequence-to-sequence model, translates text into Mel spectrograms representing speech acoustics. A WaveRNN vocoder transforms these spectrograms into corresponding audio waveforms. Finally, noise reduction algorithms refine the generated speech for enhanced clarity and naturalness. This system significantly enhanced its cloning process by training on a diverse multi-accent dataset (with 80% Indian accent). The improvement is attributed to the model being exposed to 600 hours of speech signals, encompassing the voices of 3000 speakers. This paper offers an open-source Python package https://pypi.org/project/voice-cloning/ specifically designed for professionals seeking to integrate voice cloning and speech synthesis capabilities into their projects. This package aims to generate synthetic speech that sounds like the natural voice of an individual, but it does not replace the natural human voice.

## 1. Introduction

Voice cloning is a process of creating a synthetic or artificial version of someone's voice. It involves using TTS technology and deep learning algorithms to train a computer model to mimic the speech patterns, intonation, and other unique characteristics of a specific individual's voice. The resulting synthetic voice can then be used to generate new speech that sounds like the original speaker. Voice cloning has become increasingly important due to its practical applications in various fields. For speech-impaired individuals, voice cloning can be a powerful tool that helps them communicate more effectively. People who have lost the ability to speak due to conditions such as stroke, Amyotrophic Lateral Sclerosis (ALS), or vocal cord damage can use voice cloning to generate speech that sounds like their own voice with the help of voice bank. This can be particularly important for these individuals as it helps them maintain a sense of identity and connection with others. The technology of voice cloning can be employed to generate artificial voices for individuals who are born with conditions like cerebral palsy or autism, and have never been able to speak. Through the utilization of voice cloning, such people can express their thoughts more comprehensively and communicate effectively with others. In addition, voice cloning can also save time and money by automating processes such as customer service and technical support, making it a more efficient tool.

Voice cloning is the process of creating a digital replica of a person's voice and building a TTS model. Through advanced deep learning techniques, this is accomplished by analysing and modelling a person's unique speech characteristics, such as tone, pitch, and accent. The model can be trained using a diverse set of speech datasets to generate synthetic speech that sounds similar to the voice of the original speaker. In order to generate a synthetic voice that emulates the distinctive qualities of an individual's natural voice, deep learning algorithms analyse a substantial amount of recorded speech data from that individual. Through this analysis, the algorithms learn the patterns and characteristics of the voice, including factors such as tone, pitch, and pacing that contribute to its distinctiveness. Once these features have been learned, the algorithms are able to construct a digital model of the individual's voice, which can then be utilized to produce new speech that closely resembles that of the original person. The quality and accuracy of the synthetic speech generated are determined by the calibre and quantity of the training data employed to train the deep learning algorithms. In general, the more high-quality training data available, the more accurate and natural-sounding the synthetic voice will be.

Speaker adaptation has been studied and implemented in TTS models for several decades [1], and has been incorporated into both Hidden Markov Model (HMM) based parametric TTS models and Deep Neural Network (DNN) based TTS models. In HMM-based models, speaker adaptation is often achieved using speaker-dependent Gaussian mixture models (GMMs) [2, 26, 27], which capture the unique characteristics of a particular speaker's voice. In DNN-based models, speaker adaptation is achieved through various techniques such as speaker embeddings [3] and adversarial training [4], which aim to learn a speaker-specific representation that can be used to generate personalized speech. Regarding voice cloning, TTS and voice conversion can be seen as comparable systems that utilize different inputs to create speech with a desired voice [5]. To separate the speaker content and voice information, U-net structure and the Vector Quantization (VQ) technology are utilized [6]. Voice cloning can be challenging due to low-quality recordings and noisy speech [7]. It requires technical expertise, specialized tools and software, and a significant amount of time and effort [8]. The quality of the audio source is crucial for creating an accurate clone, and the difficulty level depends on factors such as the complexity of the voice and the availability of training data. Unique accents or speech patterns may pose additional challenges.

Voice cloning technology is accent-based because it relies on the characteristics of a specific accent to create a voice that sounds like a particular individual with that accent. The role of accent in voice cloning is significant as it defines the distinct way in which people speak and their individual vocal traits. An open-source project called CorentinJ/Real-Time-Voice-Cloning was created by Corentin Jemine, a machine learning engineer [9]. The project employs deep learning technology to facilitate real-time voice cloning, allowing users to clone and synthesize their own or others voices in real-time. However, the system is trained using the LibriSpeech dataset [22], which is not tailored to Indian accents. Different accents have unique phonetic features, such as intonation patterns, stress, and rhythm [28-30]. For example, British English has different vowel sounds than American English, and speakers of different accents may place emphasis on different syllables or words [31]. To create an accurate voice clone, the network needs to be trained on recordings from speakers with the same accent as the person being cloned [11]. Therefore, this paper aims to build upon this project by developing a voice cloning system for Indian accents using a large training dataset consisting of 400 hours of Indian speech.

## 2. Related Work

The area of voice cloning focuses on generating speech that has the timbre of a new speaker, using only a small amount of speech from the new speaker [14-16]. In the context of voice cloning, two primary methodologies are employed: speaker adaptation techniques and speaker encoder techniques. Speaker adaptation involves refining the multi-speaker speech synthesis model by incorporating the speech of the new speaker. However, this approach can lead to overfitting when trained with limited data. To address this, a two-cascade module approach is suggested in [32], in which the first module predicts the acoustic characteristics of the speech, and the second module models the speaker's timbre. This approach permits the fine-tuning of only the modules associated with the speaker's timbre, which reduces overfitting. Adaspeech [33] utilizes conditional layer normalization to reduce the number of fine-tuning parameters, enabling the model to acquire knowledge of the speaker's distinct timbre without sacrificing voice quality. In addition to speaker encoder and adaptation methods, there have been advancements in neural network-driven TTS synthesis systems, i.e., blend of RNNs and CNNs. This system produces synthetic speech from textual input, capable of scaling to large datasets, handling multiple languages and accents [10].

Deep Voice 1, an advanced TTS system, is built upon a sequence-to-sequence (seq2seq) model with an attention mechanism [37]. The system comprises an encoder that converts the textual input into a continuous representation, and the decoder leverages this representation to generate the speech waveform. Deep Voice 2 improves upon Deep Voice 1 by introducing a new architecture called ClariNet [38], which combines the seq2seq model with a Generative Adversarial Network (GAN) that produces high-quality audio. Deep Voice 3 uses a Transformer architecture [20], which synthesizes speech directly from text and audio data, achieving state-of-the-art results on LJSpeech dataset and multiple languages. It employs several techniques including teacher forcing, attention mechanisms, and layer normalization to predict Mel spectrograms from text input and convert them into speech using a neural vocoder. Yi Ren et.al [12] proposed FastSpeech 2, a feed-forward Transformer architecture that enables parallel processing and significantly increases the speed of the TTS process. This approach is complementary to Deep Voice 3's sequence-to-sequence model, as both aim to improve the speed and quality of the TTS process using different techniques. Using transfer learning techniques and speaker embeddings extracted from a speaker verification system, Jiaqi Su et al. [13] developed a TTS system that synthesize speech in multiple speaker voices. When developing a TTS system targeted towards speakers with a limited quantity of training data, it is advantageous to employ a pretrained model such as Deep Voice or FastSpeech [34-36]. Tacotron2 implements the Global Style Token (GST) in conjunction with a seq2seq model to directly generate speech from textual characters. Furthermore, it incorporates an attention mechanism to enhance the quality of the speech [18]. In contrast, Attentron employs an attentive feed-forward network to predict Mel-spectrograms based on textual input. These spectrograms are subsequently converted into speech utilizing a WaveNet vocoder [39].

3. **Neural TTS System**

The proposed Neural TTS system comprises three vital models: the encoder, the synthesizer, and the vocoder. Each model plays a critical role in the process of voice cloning and speech synthesis. The encoder, which utilizes the LSTM framework, serves the purpose of speaker verification and capturing distinct voice characteristics. By verifying speakers' identities and encoding their unique vocal traits, the encoder enables precise voice cloning and speech synthesis. Operating in a seq2seq manner, the synthesizer generates Mel spectrograms from input text. This process seamlessly converts text into corresponding speech representations, ensuring high-fidelity speech synthesis. On the other hand, the vocoder transforms Mel spectrograms into waveforms in the time domain. This transformation is essential for generating natural and expressive speech output. To create a diverse and comprehensive system, all three models are trained using data from both Indian and Western accents. With a specific focus on incorporating more Indian accent data, the models underwent training with a 400-hours of Indian accent dataset.

Once these models are trained, they are assembled into a Python package named "Voice-Cloning 0.0.9." This comprehensive package enables users to engage in voice cloning and speech synthesis with ease. Notably, it offers users the unique capability to replicate both Indian and Western accents, granting a diverse array of voice cloning options. Users can input any desired text and combine it with a reference or existing voice for the desired output. Moreover, the "Voice-Cloning 0.0.9" package empowers users to develop their own personalized TTS systems. Additionally, it facilitates the replication of their own voices, allowing them to create custom speech models for a personalized touch. This package's versatility and potential make it an invaluable resource for individuals who have lost their voices due to medical conditions or surgeries. Moreover, it serves as a powerful tool for professionals seeking to integrate voice cloning or speech synthesis capabilities into their projects. The package's extensive capabilities, combined with the comprehensive training of the encoder, synthesizer, and vocoder models, ensure exceptional performance in speech synthesis and voice cloning tasks.

Section IV of the manuscript discusses the speech corpora employed in the research. In Section V, the paper outlines process involved in speech synthesis system, encompassing the training of three distinct models: the speaker encoder, synthesizer, and vocoder. In Section VI, the evaluation process which includes subjective evaluation such as MOS by human listeners and objective evaluation on unseen speakers are described. Section VII explains the process of packaging the neural TTS model and releasing the Python package.

### 4. Speech Corpora

The speech corpus chosen to train the speech synthesis system carefully covers both Indian and Western accents with 3,000 speakers representing a mix of male and female voices. This enables the system to accurately synthesize speech in both accents. The training data of the encoder model includes a diverse set of datasets, namely LibriSpeech, NPTEL2020-Indian-English-Speech-Dataset [23], LJSpeech [25], and VCTK [40]. It encompasses a wide range of dialects, such as Tamil, Odiya, Telugu, Malayalam, Bangla, as well as American, Welsh, British, and Australian accents. The overall training data of encoder covers a total duration of 600 hours, equally distributed with 480 hours dedicated to Indian accents and 120 hours to Western accents. To train the synthesizer effectively, both audio recordings and transcriptions are crucial. AccentDB [24] dataset is utilized for training the synthesizer consists of Indian accents such as Odiya, Telugu, Malayalam, and Bangla excluded Tamil accent. To address this gap, the authors developed an Indian-Tamil accent speech database, which includes speech data from 10 speakers each uttered 250 utterances. The recordings are made using a Condenser Studio XLR microphone in a laboratory environment at a sampling rate of 16 kHz. Western accents such as British, American, and Welsh from the LJ speech [25] and LibriSpeech [22] datasets are also included in the training data. The synthesizer will output generated audios and spectrograms to its model directory when training. Using this synthesized output, training data for the vocoder has been generated.

# 5. Process involved in speech synthesis system

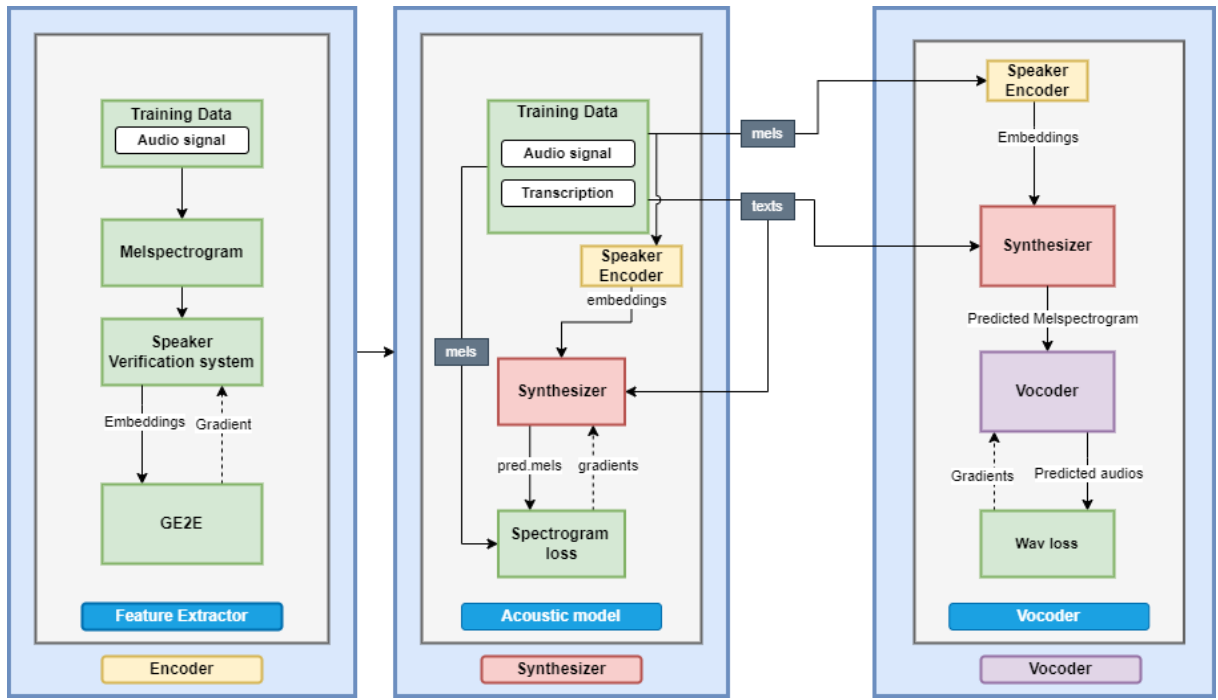

Fig.1 Architecture of speech synthesis system

Below section elaborates on the training methodology for the Speaker Encoder model, focusing on data pre-processing and the encoding of unique speaker characteristics. The Synthesizer model's training encompasses TTS model training, conditioning with speaker embeddings, and optimizing mel-spectrogram predictions. The Vocoder model is comprehensively investigated, including selecting a suitable generative model, precise mel-spectrogram and audio waveform alignment, and generation of high quality synthesized audio.

## 5.1 Speaker Verification System (Encoder)

The primary function of the speaker encoder is to capture the distinctive characteristics exhibited by different speakers. LSTM framework is utilized for text-independent speaker verification in the speaker encoder model. By processing log-mel spectrogram frames from speech utterances of varying lengths, this generates fixed-dimensional embedding vectors, commonly referred to as d-vectors. These d-vectors effectively encapsulate the unique attributes of individual speakers, enabling the system to accurately represent and differentiate speaker identities in the input data. Throughout the training process, the network is optimized to generate embeddings that exhibit a high cosine similarity for utterances originating from the same speaker. Simultaneously, the embeddings of different speakers are deliberately positioned far apart in the embedding space. The outputs of the embeddings are averaged and normalized to generate the ultimate utterance embedding, ensuring a consistent and standardized representation for each speech segment.

The encoder utilizes a training method called Generalized End-to-End (GE2E), which can handle a considerable amount of speech samples as a batch. These batches consist of N speakers, with an average of M utterances from each speaker [17]. The GE2E Loss is a loss function that helps in the training of speaker verification systems. These systems are specifically designed to identify whether two speech utterances belong to the same speaker or not. This loss considers a group of utterances from the same speaker, referred to as an "utterance group". The loss function computes the distances between the embeddings of all the utterances in the group and a weighted centroid embedding, which represents the average acoustic characteristics of the speaker. The weights assigned to each utterance in the group are learned during training and reflect the relative importance of each utterance in capturing the speaker's acoustic characteristics.

In order to create a batch, a set of $N \times M$ utterances are utilized with each utterance being spoken by a $N$ distinct speakers and each speaker contributes M utterances. Each feature vector $x_{ji}$ ($1 \leq j \leq N$ and $1 \leq i \leq M$) represents the extracted features corresponding to speaker 'j' and utterance 'i'. Then the features extracted from each utterance $x_{ji}$ are feed into an LSTM network.

The embedding vector (d-vector) is calculated by applying L2 normalization to the network output. This normalization process is mathematically expressed as Equation (1).

$$e_{ji} = (f(x_{ji}; w)) / \|f(x_{ji}; w)\|^2 \qquad (1)$$

Therefore, $e_{ji}$ denotes the embedding vector that pertains to the i$^{th}$ utterance of the j$^{th}$ speaker. The resulting output of the system can be expressed as $f(x_{ji}; w)$, where both $w$ and $x_{ji}$ correspond to the parameters of the LSTM layer and the linear layer, respectively.

The similarity matrix $S_{ji,k}$ is defined as the scaled cosine similarities between each embedding vector $e_{ji}$ to all centroids $c_k$ ($1 \leq j, k \leq N, and\ 1 \leq i \leq M$) is given in Equations (2) and (3).

$$s_{ji,k} = w \cdot cos(e_{ji}, c_k) + b \qquad (2)$$

$$c_k = E_{m[e_{km}]} = \frac{1}{M} \sum_{m=1}^{M} e_{km} \qquad (3)$$

Where 'w' and 'b' are learnable parameter and 'k' represents the index of the centroid. The LSTM's sequence-to-vector mapping produces a fixed-size embedding vector represented by 'e', while the L2 normalized response is indicated by $\{e_j \sim, e_{k1}, \ldots, e_{kM})\}$.

---

*Algorithm: Speaker verification training*

---

- Pre-processing the input speech signal to extract MFCC features.
- Using UMAP projection to reduce the dimensionality of the MFCC features.
- Feeding the projected features into a neural network.
- Training the network using the GE2E loss function for learning the speaker's characteristics.
- During training, randomly sample a batch of speech segments from different speakers and compute the GE2E loss by comparing the embeddings of all speakers in the batch.
- At inference time, taking a new spoken phrase as input.
- Using the trained system to map the features of the input to an identity vector.
- Comparing the input vector to a set of pre-defined identity vectors that corresponds to authorized speakers.
- If the input vector is similar to one of the authorized identity vectors, the speaker as valid.

---

The batch of speakers are grouped by colours to indicate different speakers based on the similarity of their speech patterns and acoustic features. The grouping of speakers improved step by step with each colour representing a different speaker in the batch. Each dot on the plot represents an embedding for a specific speech segment. By grouping the embeddings for each speaker together in the same colour represents how well the model is able to separate embeddings for different speakers as shown in Fig.2. Ideally, the embeddings for each speaker should cluster together in a tight group, while also being well-separated from embeddings for other speakers. The embeddings for each speaker are grouped and visualized using UMAP. The resulting clusters demonstrate that the embeddings capture distinct speaker characteristics, enabling speaker encoding tasks as shown in Fig. 2.

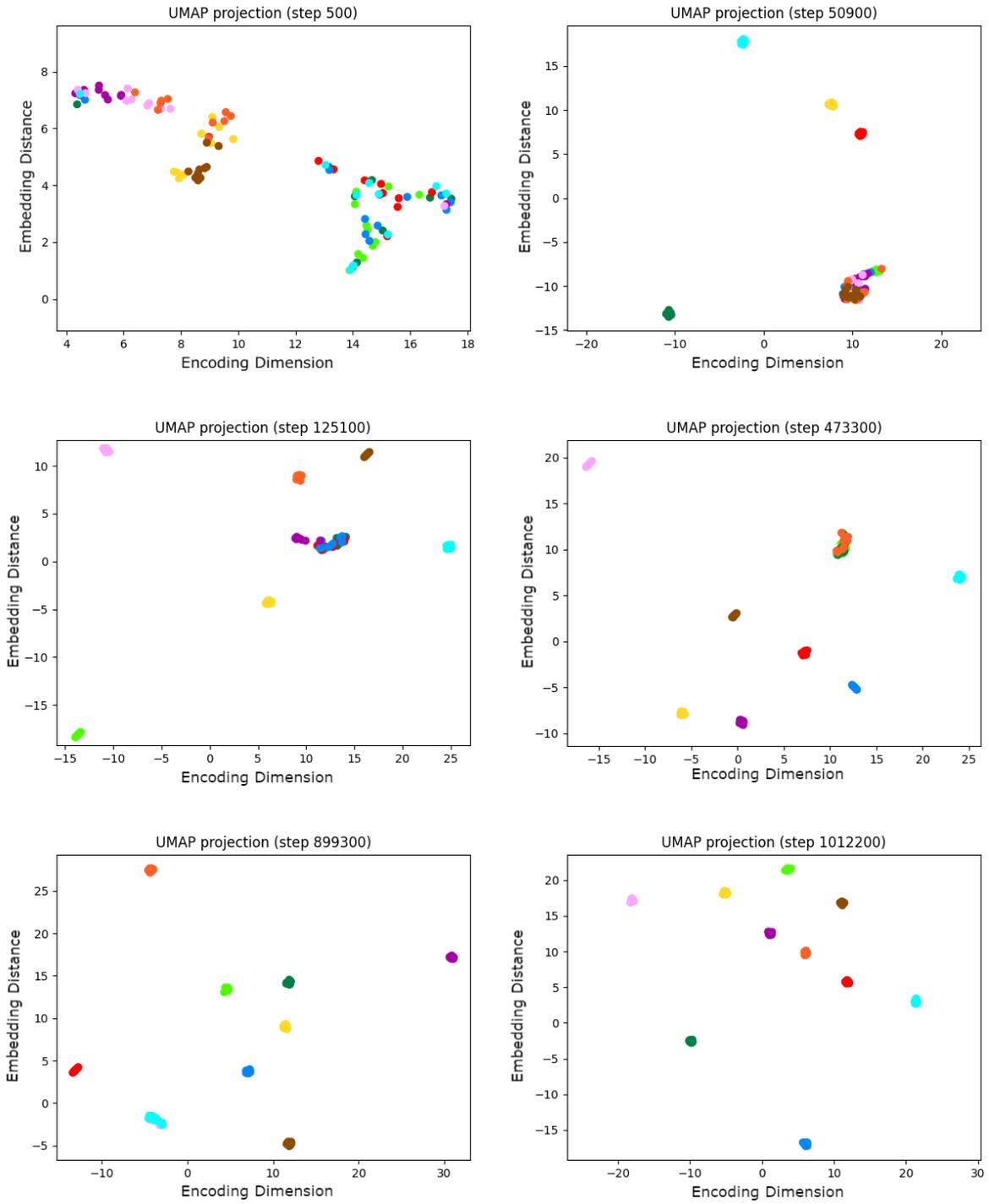

Fig.2. UMAP projection of speaker embeddings

### 5.2 Synthesizer

The synthesizer utilizes the text-to-mel model that takes textual input and predicts a corresponding Mel spectrogram. The Mel spectrogram captures the frequency content of the speech over time. To enhance the pronunciation of less common words and proper nouns, the text is transformed into a sequence of phonemes within the text-to-mel model. The system uses transfer learning with pretrained speaker encoder to extract a speaker embedding from the target audio. The target audio serves as the speaker reference signal during training. Once the text-to-mel model has predicted the Mel spectrogram, it is used to condition the WaveRNN vocoder. The WaveRNN model takes the desired Mel spectrogram representation as input and learns to predict the corresponding waveform. Finally, the WaveRNN vocoder takes the predicted Mel spectrogram as input and generates the corresponding raw audio signal. During the training process, the system utilizes pairs of text transcripts and target audio. Transfer learning is employed with a frozen, pretrained speaker encoder to extract a speaker embedding from the target audio. The target audio serves as the speaker reference signal during training. The WaveRNN model takes the target Mel spectrogram representation as input and learns to predict the corresponding waveform, resulting in the generation of the predicted Mel spectrogram. Few illustrations of the target and predicted Mel spectrograms from multiple speakers is shown in Fig. 3.

| *Algorithm: Synthesizer training* |
|---|
| - Converting the input text transcript to phonemes or graphemes using a text normalization method. |
| - Generating a sequence of phoneme or grapheme embeddings using an encoder. |
| - Transmitting the embeddings through a pre-net to reduce the dimensionality and improve conditioning. |
| - Feeding the pre-net output through a text-to-mel model, which predicts a mel-spectrogram representation of the speech signal. |
| - Passing the predicted mel-spectrogram through the WaveRNN model, which generates the corresponding raw audio signal. |

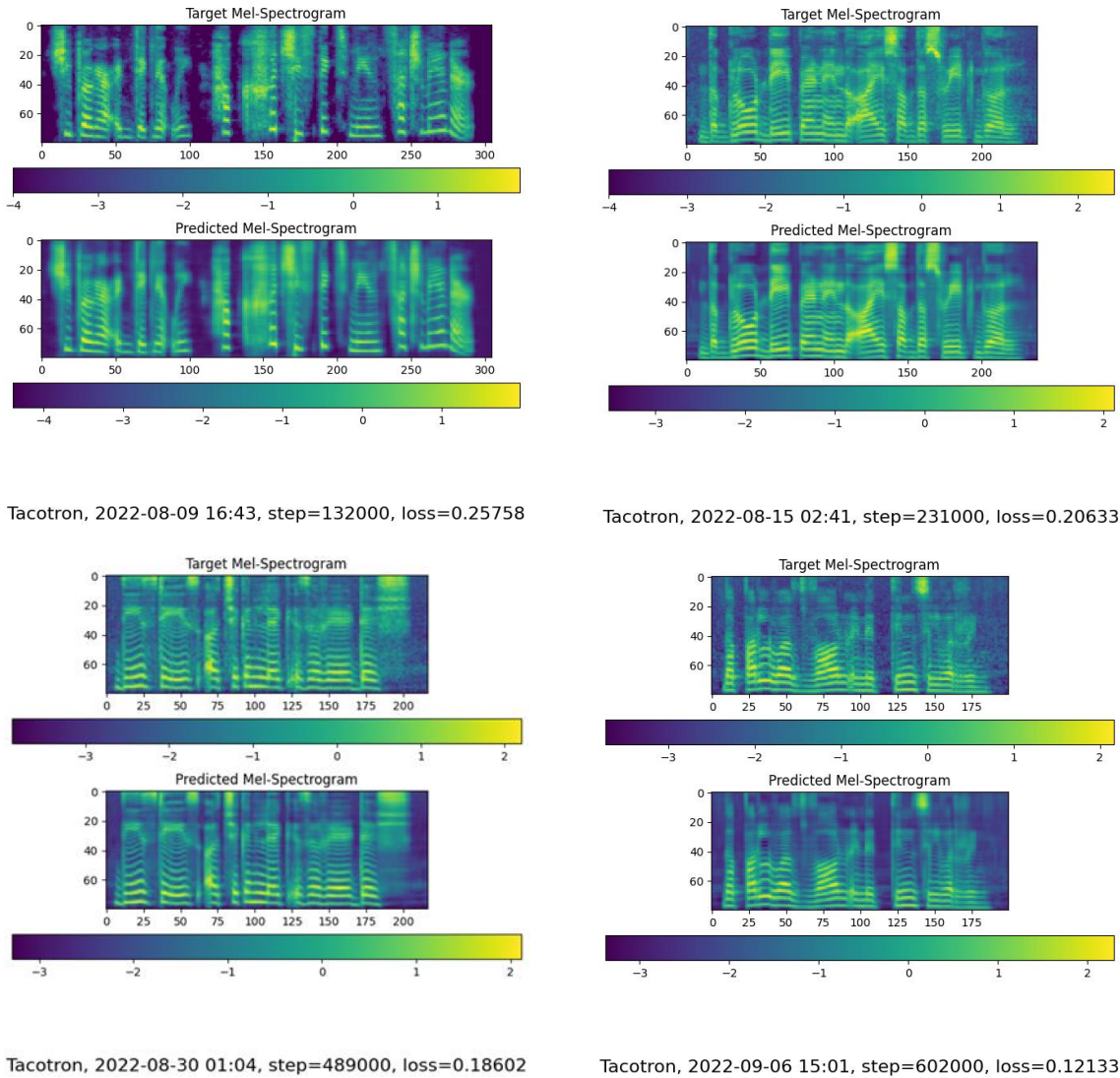

Fig.3. Comparison of Target and Predicted Mel Spectrograms for Multi- Speaker Synthesis

**5.3 Vocoder**

The speaker vocoder is used to generate high-quality audio waveforms directly from the digitized raw audio using autoregressive WaveRNN architecture. This task is challenging due to the complexity and high dimensionality of raw audio signals. The autoregressive WaveRNN architecture takes Mel spectrogram as input to generate high-quality speech that resembles the voices of various individuals. To create a Mel spectrogram from a raw audio signal, a Fourier transform is applied to obtain the frequency content of the signal. The resulting spectrum is then filtered using a filter bank to produce a set of filter bank outputs. These outputs are then transformed logarithmically to generate the Mel spectrogram. Then the synthesized Mel spectrogram is converted into time-domain waveform samples by modelling the conditional distribution of the next waveform sample, given the previous samples and the Mel spectrogram input. This enables the network to generate realistic and high-quality speech that clone the voices of various individuals, resulting in a natural-sounding output. To create a multi-speaker vocoder, the network is trained on data from many different speakers.

To train the vocoder, the initial step involves pre-processing the training data to acquire the corresponding Mel spectrograms. Subsequently, these Mel spectrograms serve as the input to the neural vocoder for further training and processing. WaveRNN architecture is used because of its effectiveness for modelling the complex patterns in raw audio signals. In WaveRNN architecture, the system is conditioned on the input Mel spectrogram by concatenating it to the input waveform at each layer of the network. The system is then trained to predict the corresponding waveform given the Mel spectrogram as the input. To train the system, a dataset of speech recordings and corresponding Mel spectrograms is used. The system is trained using a Mean Squared Error (MSE) loss function to quantify the disparity between the predicted and target waveforms. To minimize this loss, a Stochastic Gradient Descent (SGD) optimizer is employed. During training, a technique called teacher forcing is used, which involves feeding the correct waveform as input to the system during each time step. Once the system has been trained, it is used to synthesize speech from new Mel spectrograms by feeding them into the system and generating the corresponding audio signal. After obtaining the synthesized speech/cloned voice, a process called noise reduce [21] is applied to remove unwanted noise and make the speech clearer.

**Algorithm: Vocoder training**

- Obtaining the training data from synthesizer output generated audios and corresponding Mel spectrograms
- Using a WaveRNN architecture, input Mel spectrogram is conditioned by concatenating it to the input waveform at each layer of the network.
- Training the network using the following steps:
    a. Initializing the network weights.
    b. Feeding the Mel spectrogram and waveform as inputs to the network using teacher forcing method.
    c. Computing the disparity between the predicted and actual waveforms by using the MSE loss function.
    d. Using an optimizer such as SGD to minimize the loss.

> e. Repeating steps b-d for a predetermined number of epochs.
- Using the trained neural vocoder to synthesize speech from new Mel spectrograms by feeding them into the network and generating the corresponding audio signal
- Applying noise reduction technique to remove unwanted noise and improve the overall quality of the speech.

## 6. Evaluation Metrics

To evaluate the effectiveness of voice cloning and speech synthesis system, three evaluation metrics are utilized: MOS is an important tool for evaluating the effectiveness of speech synthesis systems, as it provides a measure of speech perceived as high-quality and natural-sounding by human listeners. GPE is used to evaluate the accuracy of speech synthesis systems by measuring the disparity between the pitch of the input signal and synthesized signal. SD measures the spectral difference between the original speech signal and the synthesized speech signal. The evaluation of the synthesized speech involves subjective measures such as MOS, as well as objective measures such as GPE and SD. The objective evaluation are applied on unseen speakers (Excluded from training), having Western accents (VCTK and LibriSpeech) and Indian accents (Tamil, Bangla, Telugu, and Malayalam).

### 6.1 Speaker similarity evaluation by human listeners

In this interface, listeners are presented with original and synthesized audio samples as shown in Fig 4 (a). The listeners are requested to rate the similarity of the speaker's voices on a rating scale that ranges from "not at all similar" to "slightly similar," "moderately similar," and "very similar. MOS is subsequently calculated by aggregating the ratings provided by all participants. The responses of the participants are recorded and visualized in Fig. 4 (b), where the "very similar" category corresponds to MOS values between 5 and 4, "moderately similar" corresponds to MOS values between 4 and 3, "slightly similar" corresponds to MOS values between 3 and 2, and "not at all similar" includes MOS values less than 2. The subjective speaker similarity MOS evaluation, conducted by human listeners holds significant importance in the assessment of speech synthesis system.

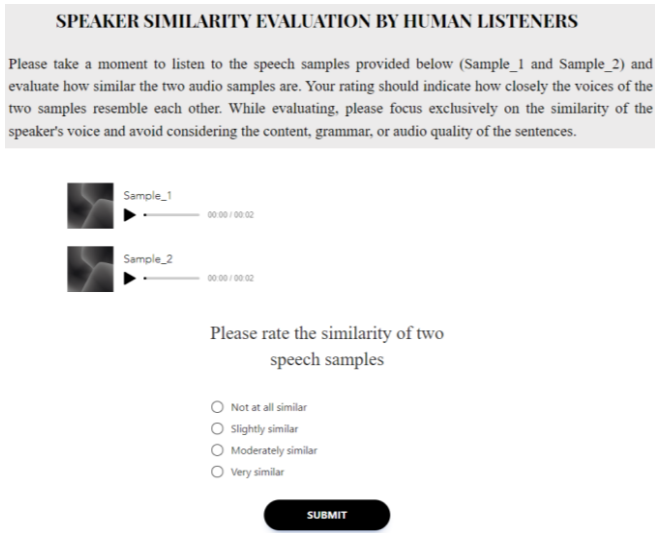

(a)                                                                 (b)

Fig. 4 (a) illustrates the MOS evaluation interface, while Fig. 4 (b) visually represents the responses provided by human listeners.

**Table. 1 Objective evaluation on unseen speakers. (A downward decreasing arrow signifies that as the metric value decreases, the performance improves)**

| Accent | Dataset | Speaker_ID | Gender | Existing work [13] | Proposed work | | |
|---|---|---|---|---|---|---|---|
| | | | | MOS | MOS | GPE ↓ | SD ↓ |
| Western | VCTK [32] | p230 | F | 4.65 | 4.64 | 1.95 | 3.38 |
| | | P240 | F | 4.67 | 4.69 | 1.69 | 2.25 |
| | | P300 | F | 4.87 | 4.76 | 2.23 | 1.43 |
| | | P340 | M | 4.71 | 4.78 | 1.23 | 2.23 |
| | | P260 | M | 4.31 | 4.58 | 1.54 | 2.67 |
| | | P270 | M | 4.77 | 4.82 | 2.33 | 1.32 |
| | LibriSpeech [22] | 7021 | M | 4.55 | 4.52 | 0.43 | 2.92 |
| | | 7729 | M | 4.48 | 4.73 | 3.89 | 1.04 |
| | | 8230 | M | 4.70 | 4.79 | 0.67 | 0.78 |
| | | 3575 | F | 4.36 | 4.45 | 1.98 | 2.67 |
| | | 4970 | F | 4.16 | 4.12 | 0.34 | 2.78 |

| | | 4992 | F | 3.81 | 4.46 | 2.12 | 3.23 |
|---|---|---|---|---|---|---|---|
| Indian | Tamil accent dataset by authors | Tamil_02 | F | - | 4.23 | 0.51 | 1.45 |
| | Accent DB [24] | Bangla_01 | F | - | 4.47 | 1.97 | 5.32 |
| | | Telugu_02 | M | - | 4.26 | 1.38 | 4.42 |
| | | Malayalam_03 | F | - | 4.17 | 1.55 | 3.79 |

**6.2 Comparison with existing work**

The MOS of the proposed work are compared with existing work as mentioned in Table 1. However, there was no pertinent literature identified for a comparison of the GPE and SD metrics. While the existing work solely relied on subjective evaluation, the proposed approach goes a step further and incorporates objective evaluations such as GPE and SD for both Western and Indian accents. Conducting the evaluation on unseen speakers enables a thorough analysis of the synthesized speech quality across a diverse range of accents and speakers. The proposed work demonstrates superior MOS score when compared with existing work [13]. A higher MOS score indicates that the synthesized speech closely resembles the original speech, which is desirable for a good voice cloning system. A low GPE value indicates that the speech synthesis system has accurately replicates the pitch of the original speech signal. Similarly, a low value for SD indicates a close match between the spectral characteristics of the synthesized speech and the original speech, suggesting a high level of accuracy in the cloning process. Overall, the table suggests that the proposed speech synthesis system is able to produce synthesized speech that closely matches the original speech, indicating a good level of performance.

**7. Release of Voice-Cloning 0.0.9 python package**

The trained neural system has demonstrated promising outcomes through both subjective and objective evaluations. As a result, the trained encoder, synthesizer, and vocoder models have been integrated into a Python package. By simply installing the voice cloning package, users gain the ability to effortlessly synthesize speech. This package holds the potential to be seamlessly integrated into assistive technologies, offering a transformative solution for individuals with speech disorders. By harnessing the power of voice banks, this technology enables individuals to restore a semblance of their natural voice. Moreover, this innovation is poised to be an helpful asset for professionals immersed in the field of speech synthesis technology. Voice-Cloning 0.0.9 python package [https://pypi.org/project/Voice-Cloning/](https://pypi.org/project/Voice-Cloning/) has been released, which empowers users to generate audio from text, design their own text-to-speech systems, and even create a personalized speech model by cloning their own voice. The entire Voice-Cloning package is conveniently available via the Python Package Index (PyPI), making installation effortless. Simply execute the following command to install Voice-Cloning package using pip:

- Use the package manager [pip](#) to install Voice-Cloning 0.0.9
- pip install voice-cloning

The availability of the package on PyPI ensures that the user can easily incorporate Voice-Cloning into their projects, enabling them to harness its powerful speech synthesis and voice cloning features without hassle. To reproduce the experiments and explore the functionalities offered by Voice-Cloning, users can refer to the comprehensive documentation provided alongside the package. This documentation explains the usage of functions, parameters, and examples, facilitating a smooth and effective utilization of the package's capabilities. Fig. 5 illustrates the functionality of the voice cloning package, which accepts any text input and a reference voice or existing voice. Fig. 6 explains the flow of speech synthesis model used in the package. Giving users the option to select between cloning via an external reference voice or speech synthesis using an existing voice from the sound library. The sound library features a collection of 31 speaker voices supporting both "Western" and "Indian" accents. The package tasks are summarized in Table 2.

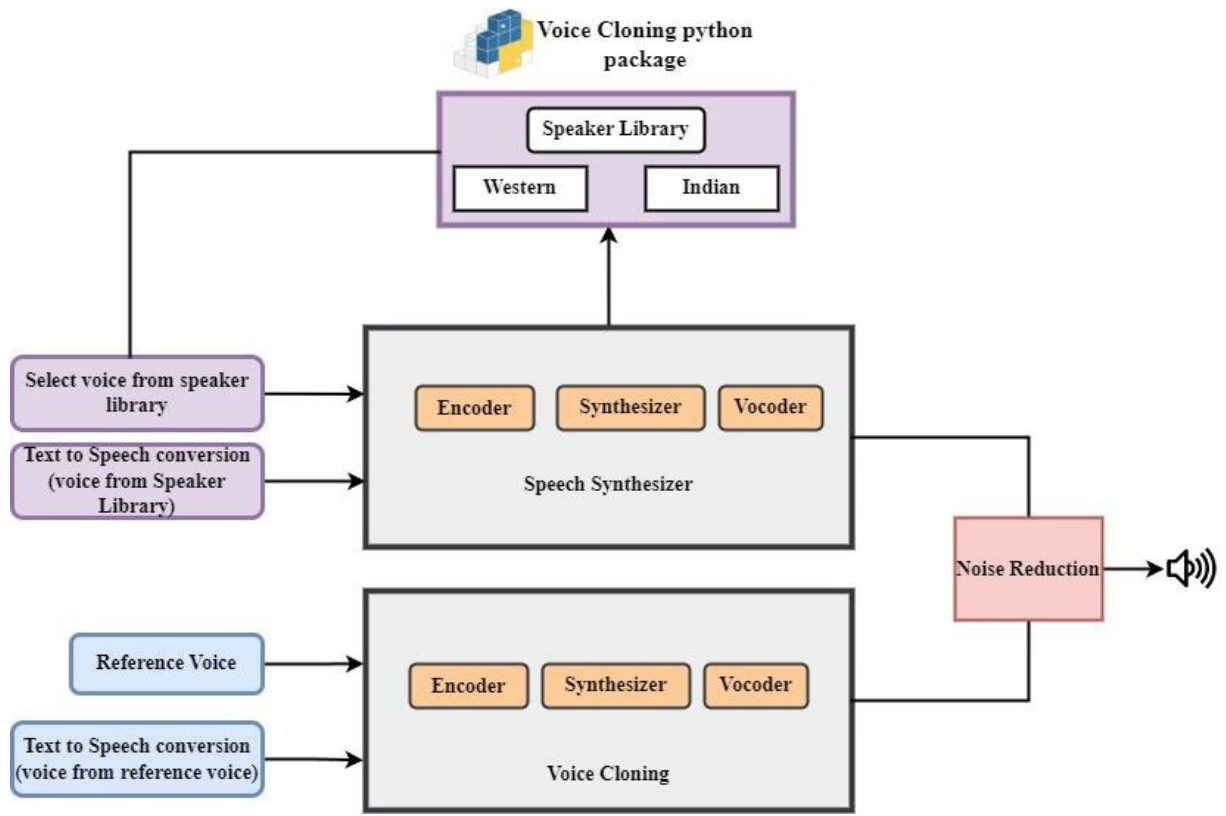

Fig.5. Pipeline of speech synthesis and voice cloning model

**Table 2. Voice-Cloning tasks**

| *Feature* | *Description* |
|---|---|
| *Speech synthesis* | Users can utilize the package to generate synthesized speech by entering text, with access to pre-loaded speakers, similar to TTS system. |
| *Real-time voice cloning* | The package offers the capability to perform real-time voice cloning by analysing either a reference voice clip or the user's speech input. |
| *Multi-Accent support* | Supports Western and Indian accents for speech synthesis and voice cloning. |
| *Noise reduction* | The package incorporates noise reduction to reduce noise in recorded audio, resulting in an overall improvement in the quality of the synthesized or cloned speech. |

## 8. Conclusion

A speech synthesis and voice cloning package has been developed, consisting of Speaker verification system, a synthesizer, and a vocoder. These three systems are trained using huge amount of open source dataset which includes accents like American, Australian, Welsh, Tamil, Telugu, Malayalam, Odiya, Bangla. The speaker encoder uses a LSTM framework to capture the unique characteristics of different speakers and conditioning the synthesis network on the desired target speaker's reference speech signal. Synthesizer consists of two main parts: the text-to-mel model and the WaveRNN vocoder. The text-to-mel model takes the text as input and predicts a corresponding Mel spectrogram, which captures the frequency content of the speech over time. The WaveRNN vocoder then uses this predicted Mel spectrogram to generate the corresponding raw audio signal. Vocoder trained on data from many different speakers i.e. the multispeaker vocoder learns to generate high-quality audio waveforms directly from raw audio signals that sounds like different people, making it a valuable tool in various fields such as speech synthesis and voice cloning. Noise reduce algorithm have been incorporated to further enhance the quality of the output audio signal. The MOS subjective evaluation involves assessing the ground truth and synthesized speech (proposed model) for unseen speakers of both Western and Indian accents. The lower the GPE score, the better the accuracy of the speech synthesis system in reproducing the pitch of the original speech signal. A low SD value indicates that the synthesized speech closely matches the spectral characteristics of the original speech, which indicates a high level of accuracy in the cloning process. However, voice cloning systems encounter challenges when attempting to reproduce the intricate emotional cues present in human speech, resulting in synthesized voices. Hence the package employs speaker encoder, synthesizer and vocoder to produce synthesized speech that closely resembles a person's natural voice and providing a support for those with speech disorders to communicate more effectively. This technology also opens up new possibilities for professionals seeking to integrate voice cloning or speech synthesis capabilities into their projects. Future work in voice cloning could explore enhancing the expressiveness and emotional range of synthesized voices, enabling them to convey a wider spectrum of human feelings and tones.

**Declarations**

**Conflict of interest**

The corresponding author on behalf of all authors, affirms that there are no conflicts of interest to disclose in relation to this work.

**Acknowledgment**

The authors express their sincere gratitude to the Science for Equity Empowerment and Development Division (SEED) under the Department of Science and Technology (DST), Government of India, for their financial support towards this project (Ref No: SEED/TIDE/2019/431/G). The team extends their sincere gratitude to Karunya Institute of Technology and Sciences, Coimbatore, Tamil Nadu, for providing the research facilities and the infrastructure.